\pdfoutput=1
\documentclass[11pt]{article}
\usepackage[utf8]{inputenc}

\usepackage{graphicx} 
\usepackage{caption}
\usepackage{subcaption}
\usepackage{booktabs}
\usepackage{listings}
\usepackage{xcolor}
\usepackage{algorithm}
\usepackage{algorithmic}
\usepackage{multirow}
\usepackage{enumitem}
\usepackage{fontawesome}
\usepackage{pifont}
\usepackage{MnSymbol}
\usepackage{indentfirst}
\usepackage{threeparttable}
\usepackage{siunitx}
\usepackage{amsfonts}
\usepackage{natbib}
\usepackage{makecell}
\usepackage{geometry} 
\usepackage{lineno}

\title{Efficient Estimation of Temporal Exceeding Probability for Ship Responses in Broadband Wave Fields}

\author{{Shayesteh Hafezi, Xianliang Gong, Yulin Pan}\\
\textit{Department of Naval Architecture and Marine Engineering} \\
\textit{University of Michigan}\\
Ann Arbor, MI, USA}
\date{}

\begin{document}
\maketitle


\begin{abstract}

In this paper, we develop an efficient method to evaluate the temporal exceeding probability of ship motion (percentage of time for the ship motion to be above a given high threshold) in an irregular wave field. Our method builds on our previous work \cite{gong2022efficient} which converts the calculation into a sampling problem in the space of wave group parameters, within which an acquisition-based sequential sampling method is developed to reduce the number of required samples. Two critical advancements are achieved in this paper relative to \cite{gong2022efficient}. (1) We develop a new wave group parameterization method, which allows the framework to be applied to general broadband wave fields. (2) We incorporate the variability regarding each parameterized wave group (e.g. varying wave form and initial condition of the ship encountering the group) into the final estimation of a single value of the temporal exceeding probability. Our complete framework is tested for a ship subject to a wave field with a JONSWAP spectrum, for different ship motion dynamical models, spectral bandwidths, and exceeding thresholds. The results show that for most cases our method provides a result with O(\SI{15}{\percent}) error or below within \num{210} samples, with the ground truth obtained from a continuous simulation that is more than \num{2300} times more expensive than our method. We also demonstrate the benefits of sequential sampling (with an acquisition function updated due to (2)) compared to standard random or Latin hypercube (LH) samplings, in terms of the mean error of the results. 

\end{abstract}


\section{Introduction}

The assessment of rare events in complex dynamical systems is crucial for reliability and safety in engineering design \cite{mohamad2018sequential}. In the marine context, one of such problems is the extreme ship motion induced by adverse sea conditions, such as parametric rolls, which poses challenge in the design and operation of ships \cite{mohamad2017direct, spyrou2008problems, umeda2004nonlinear}. These rare events, although occurring infrequently, can have catastrophic consequences, necessitating accurate estimation of their probabilities for informed design and decision-making processes. 

Estimation of the statistics of wave-induced extreme motions is a challenging problem due to the stochastic nature of the wave field and the low probability of these extreme events. A brute-force calculation using a high-fidelity method (such as CFD) is usually computationally prohibitive, since a very long simulation of ship-wave interaction is needed to obtain sufficient exposure of the extreme events \cite{belenky2012approaches}. Over the years, many reduced‐order methods have been developed to enable the calculation, such as the Envelope POT approach \cite{mctaggart2000ship,campbell2016application,weems2019envelope}, the split‐time approach \cite{belenky2012approaches,belenky2016split,belenky2018tail}, and the critical‐wave‐group approach \cite{themelis2008probabilistic,anastopoulos2016ship,anastopoulos2019evaluation,anastopoulos2016towards}.
In spite of their success in reducing the computational cost, heuristic components are more or less involved which affects the robustness of their probability estimation in realistic situations. 

More recently, a new group of methods have been proposed for estimation of the extreme ship motion probability \cite{gong2022efficient, mohamad2018sequential, cousins2016reduced, farazmand2017reduced, gong2021full, gong2022sequential, tang2022estimating, gong2024multifidelity, gong2022effects}. These methods convert the probability estimation problem into a sampling problem in a low-dimensional parameter space, which describes the encountering wave condition associated with input probability. A sequential sampling algorithm based on a Gaussian process regression (GPR) surrogate model is then developed to reduce the number of necessary samples (each associated with a ship motion simulation) for estimation of extreme ship motion probability. Following the above framework, such methods have been first developed for the group-based motion probability \cite{mohamad2018sequential, gong2021full, tang2022estimating} (i.e., distribution of largest motion within a group), and then for temporal exceeding probability (i.e., percentage of time for the ship motion to be above a given high threshold) \cite{gong2022efficient}. As discussed in \cite{gong2022efficient}, the temporal exceeding probability, which we will continue to explore in this paper, provides a more robust and general probabilistic description of extreme motions than the group-based counterpart. While this group of methods are developed on a more rigorous foundation than previous ones, their applications are mostly restricted to narrow-band wave fields. This limitation stems from the wave field parameterization: as shown in Figure~\ref{fig:narrow_band}, a narrow-band wave field consists of consecutive Gaussian-like wave groups that can be naturally parameterized by the group length $l$ and height $a$. However, a broadband wave field lacks such structures since more frequency components (other than simply a modulation of the main frequency) are involved (see Figure~\ref{fig:group_parameterization}).

One work that deals with the broadband wave field is \cite{guth2022wave}, which parameterizes the wave field using Proper Orthogonal Decomposition (POD) instead of the Gaussian wave group approach. Although POD naturally captures the structure of a broadband wave field, the dimension of the parameter space increases up to \num{20}-\num{30} compared to only \num{2} ($l$ and $a$) in the narrow-band cases. For such a high-dimensional parameter space, sequential sampling is more difficult (or at least significantly more expensive) and is thus not implemented in \cite{guth2022wave}. In addition, the number of dimensions needed for the input parameter space depends on the spectral bandwidth, with the relation not clearly known \emph{a priori}.    

In this paper, we extend our methodology in \cite{gong2022efficient} to enable the evaluation of temporal exceeding probability in broadband wave fields. A key feature of the developed method is that we are able to retain a two-dimensional parameterization of the wave field (in contrast to up to 20-30 dimensions as in \cite{guth2022wave}). This unique feature has been made possible due to two critical developments: (a) we apply a new wave parameterization method inspired by \cite{bassler2010application}, which parameterizes the wave field by length $l$ and height $a$ of wave groups exceeding a pre-selected threshold. (b) We approximate the randomness of the response due to variability associated with each $(l, a)$ group by a Gaussian distribution, and incorporate its effect into the final estimation of the temporal exceeding probability. Here the ``variability'' includes detailed wave form and initial condition of a ship encountering this wave group, the former associated more with the broadband wave field, and the latter general for wave fields of any spectral width but not handled in \cite{gong2022efficient}. A sequential sampling method is correspondingly developed considering (b), in particular in terms of tracking the randomness of the response function as a part of the posterior of the GPR. We note that in the whole process, the group threshold in (a) serves as a free parameter whose optimal value is not known \emph{a priori}. Our extensive tests show that a choice of it close to half of the significant wave height $H_s$ produces satisfactory results for all cases.

We test our method for broadband wave fields with JONSWAP spectra (characterized by small peak enhancement parameters), with the ship response calculated from a nonlinear roll model. Different spectral bandwidths, exceeding thresholds, and parameters in the roll model are chosen, resulting in a total of 36 test cases. We show that, in most cases, our method achieves a relative error of O(\SI{15}{\percent}) or below within \num{210} samples, compared to the ground truth obtained from one continuous simulation that is typically over \num{2300} times more expensive. Moreover, we illustrate the advantages of the developed sequential sampling method over traditional random and Latin hypercube (LH) sampling methods, in terms of faster convergence of the result with the number of samples. We finally remark that the coupling of the current method with high-fidelity ship motion simulations (such as CFD) is straightforward, as has been demonstrated in \cite{gong2022efficient}. This will not be demonstrated again in the current paper.     

The remainder of this paper is organized as follows. In Section 2, we provide the problem setup on temporal exceeding probability. In Section 3, we present the methodology, especially the new wave group parameterization and the sequential sampling framework. The validation of the method is demonstrated in Section 4, with Section 5 concluding the paper. 

\section{Problem setup}
Considering a ship navigating through a wave field with wave-induced responses, say roll motions, we are interested in evaluating the temporal exceeding probability given by: 
\begin{equation}
P_{\text{temp}} = \frac{1}{T_{\text{end}}} \int_{0}^{T_{\text{end}}} \mathbf{1}(|r(t)| - r_s) \, \mathrm{d}t,
\label{eq:P_temp}
\end{equation}
where $r(t)$ is the time series of roll angle resulted from wave signal $\eta(t)$, $r_s$ is a high threshold above which the roll is considered extreme, $t$ is time and $T_\text{end}$ is the total exposure time,  and \( \mathbf{1}(\cdot) \) is the Heaviside (indicator) function defined as:
\begin{equation}
\mathbf{1}(x) = \begin{cases} 1, & x > 0 \\ 0, & x \leq 0 \end{cases}.
\end{equation}
Therefore, Eq.~\eqref{eq:P_temp} describes the percentage of time that the ship roll response exceeds a high threshold given sufficient exposure time. 

A direct evaluation of Eq.~\eqref{eq:P_temp} involves a computation of roll motion $r(t)$ from wave signal $\eta(t)$ in time $[0,T_\text{end}]$. This can become computationally prohibitive if high-fidelity simulations (such as CFD) are employed, since $T_\text{end}$ needs to be very large to cover enough exceeding events associated with a given wave spectrum. Our goal is to develop an efficient method (free from the long computation of $r(t)$) to enable the evaluation of Eq.~\eqref{eq:P_temp} in general wave fields including those with broadband spectra.  
\section{Method}

\subsection{Review of the previous method in \cite{gong2022efficient}}

\begin{figure}[htbp]
\centering

\begin{subfigure}{0.5\textwidth}
    \includegraphics[width=\textwidth]{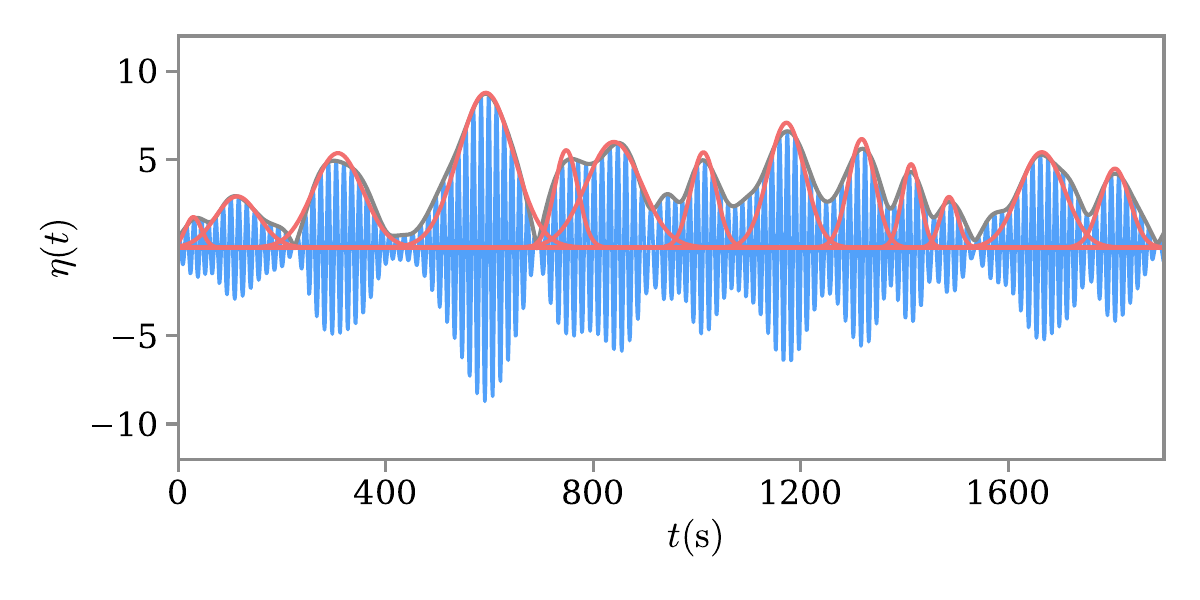}
    \vspace{-2em}
    \caption{}
    \label{fig:narrow_with_l_a}
\end{subfigure}
\begin{subfigure}{0.3125\textwidth}
    \includegraphics[width=\textwidth]{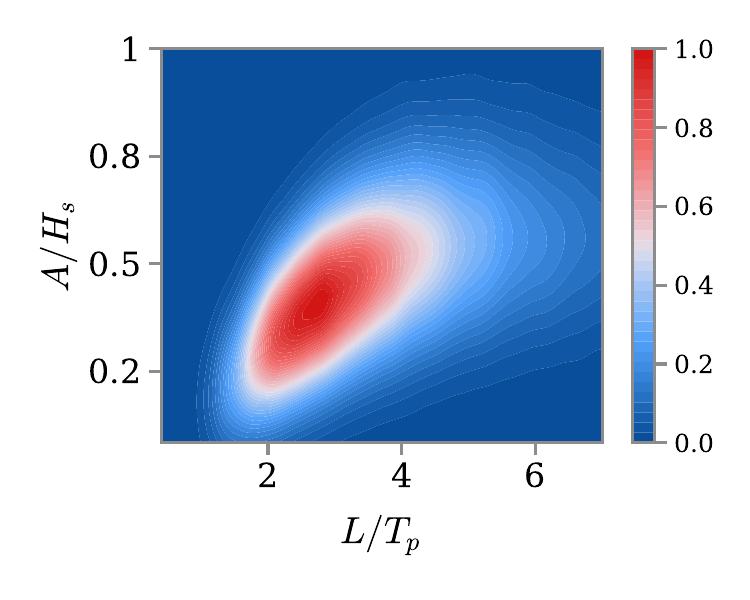}
    \vspace{-2em}
    \caption{}
    \label{fig:group_P_LA_narrow}
\end{subfigure}

\definecolor{BLUE}{rgb}{0.32, 0.63, 0.98}
\definecolor{PINK}{rgb}{0.95, 0.44, 0.44}
\definecolor{GRAY}{rgb}{0.54, 0.54, 0.54}

\caption{(a) Surface elevation $\eta(t)$ (\textcolor{BLUE}{\rule[0.5ex]{1.5em}{1pt}}) and the corresponding envelope process (\textcolor{GRAY}{\rule[0.5ex]{1.5em}{1pt}}) fitted by an ensemble of Gaussian wave groups (\textcolor{PINK}{\rule[0.5ex]{1.5em}{1pt}}) with parameters $L$ and $A$ in a narrow-band wave field for a typical Gaussian spectrum. (b) $P_{LA}(l, a)$ obtained from the wave field with spectral peak period of $T_p= \SI{15}{s}$ and the significant wave height of $H_s=\SI{12}{m}$.}
\label{fig:narrow_band}
\end{figure}

In \cite{gong2022efficient}, we have developed a method to evaluate Eq.~\eqref{eq:P_temp} for narrow-band wave fields with some other simplifications. The method starts by describing the wave elevation $\eta(t)$ by consecutive Gaussian groups parameterized by length $l$ and height $a$ (see Figure~\ref{fig:narrow_with_l_a}), a procedure only appropriate for waves with a narrow-band spectrum. By collecting $(l, a)$ from all $m$ groups in the wave time series $\eta(t)$, a probability distribution $P_{LA}(l,a)$ can be generated (see Figure~\ref{fig:group_P_LA_narrow}). The temporal exceeding probability can then be (approximately) computed by:
\begin{equation}
P_{\mathrm{temp}}^{a} \;=\; \frac{1}{\,T_{\mathrm{end}}\,} \sum_{i=1}^{m} S\bigl(l_{i},a_{i}\bigr)
\;=\;\frac{m}{\,T_{\mathrm{end}}\,}\,
\int\;S(l,a)\;\,P_{LA}(l,a)\,\mathrm{d}l\,\mathrm{d}a,
\label{eq:ptemp_approx_narrow}
\end{equation}
where
\begin{equation}
    S\bigl(l_{i}, a_{i}\bigr)\;=\;\int\; \mathbf{1}\!\bigl(\bigl|\,r\bigl(t; l_{i}, a_{i}\bigr)\bigr| \;-\; r_{s}\bigr)\,\mathrm{d}t,
    \label{eq:S_narrow}
\end{equation}
is the exceeding time of the $(l_i, a_i)$ group, so that the integral in Eq.~\eqref{eq:S_narrow} gives the expected exceeding time in one wave group.

Clearly, in the computation of Eq.~\eqref{eq:ptemp_approx_narrow}, the key is to obtain the function $S(l, a)$. Since a direct computation of $S(l, a)$ for all groups can be prohibitively expensive, a GPR (the details of the algorithm and formulae are summarized in Appendix~\ref{appendix:first}), in particular the mean value, is used as a surrogate of

\begin{equation}
    S(l,a) \sim \text{GP}\left( \mu_{S}(l, a), \sigma^2_{S}(l, a ) \right),
    \label{eq:S_before}
\end{equation}
trained through data from limited number of samples in $(l, a)$ space with associated function evaluations. In order to reduce the number of required samples, a sequential sampling algorithm based on Bayesian experimental design (BED) method is developed, which samples at optimal locations to accelerate the convergence of the results to the true $P_{\text{temp}}^a$.

We note that $P_{\text{temp}}^a$ in Eq.~\eqref{eq:ptemp_approx_narrow} serves only as an approximation to $P_{\text{temp}}$ in Eq.~\eqref{eq:P_temp} (thus having the superscript $a$ in the former). This is because when using Eq.~\eqref{eq:ptemp_approx_narrow}, one has to assume that the group exceeding time is deterministic given $(l, a)$, i.e., $S(l, a)$ is a deterministic function. However, in practice given $(l, a)$, the group exceeding time can vary because both ship initial condition encountering a wave group and finer details of the wave elevation in a group can be different. The former is a problem for any wave field, and the latter becomes more severe for those with a broadband spectrum (of course after a proper group parameterization is developed as will be discussed). A new approach is needed to address this variability problem regarding $S(l, a)$.

\subsection{The proposed method}
\label{sec:method}

We now introduce our new approach, which is applicable to wave fields with a broadband spectrum and addresses the variability problem regarding $S$. 

\begin{figure}[htb!]
\centering
\begin{subfigure}{0.5\textwidth}
    \includegraphics[width=\linewidth]{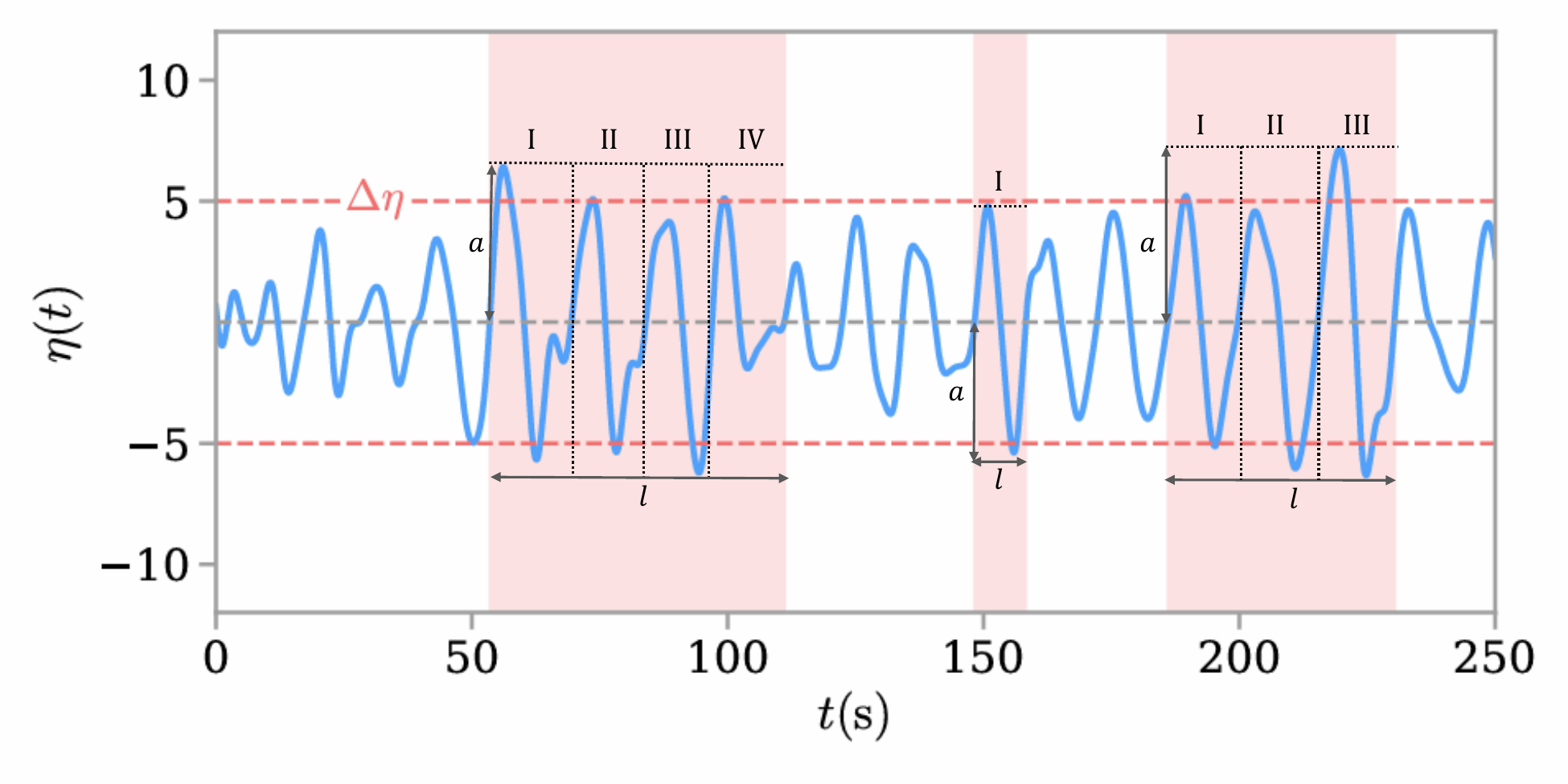}
    \vspace{-2em}
    \caption{}
    \label{fig:broad_group_param}
\end{subfigure}
\begin{subfigure}{0.3125\textwidth}
    \includegraphics[width=\linewidth]{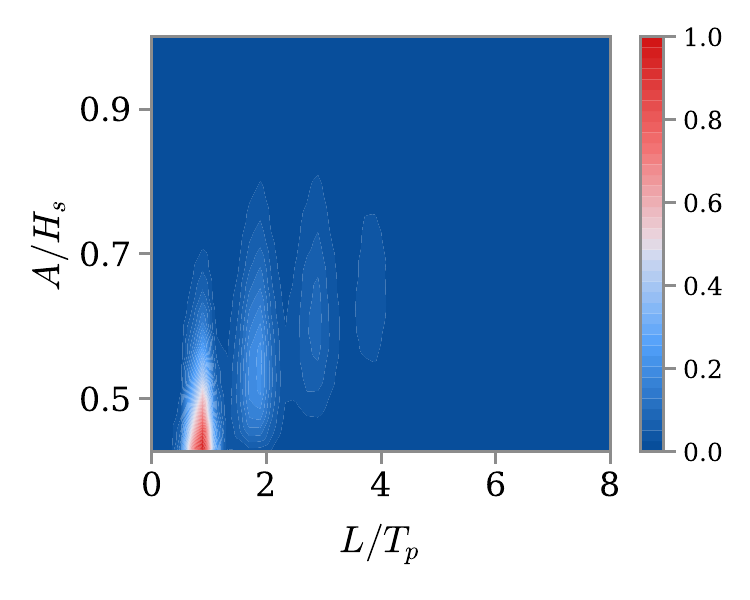} 
    \vspace{-2em}
    \caption{}
    \label{fig:group_P_LA_broad}
\end{subfigure}

\caption{(a) An example of wave group parameterization in a broadband wave field with a JONSWAP spectrum with peak enhancement factor $\gamma=3$. Three groups (marked by the shaded areas) are identified in this wave segment according to $\Delta \eta =\SI{5}{m}$. The three groups are associated with 4, 1, and 3 single waves, respectively, with each one characterized by its own values of $(l,a)$. (b) Probability distribution \( P_{LA}(l, a) \) calculated from collections of wave groups, with $L$ and $A$ normalized by the spectral peak period $T_p= \SI{15}{s}$ and the significant wave height $H_s=\SI{12}{m}$, respectively.}
\label{fig:group_parameterization}
\end{figure}

Inspired by \cite{bassler2010application}, a new group parameterization method is developed which targets the critical segments in $\eta(t)$ that may lead to extreme ship responses. Specifically, we first define a single wave as a segment starting from $\eta=0$ with a positive slope and experiencing two subsequent zero-crossings. A wave group is then defined as a series of consecutive single waves with amplitudes above a given threshold $\Delta \eta$ (see Figure~\ref{fig:broad_group_param}). Each wave group is then parameterized by its length $l$ and maximum amplitude $a$ within the group, with the collection of all groups providing a two-dimensional parameter space with known probability density $P_{LA}(l, a)$ (Figure~\ref{fig:group_P_LA_broad}).

We note that the threshold $\Delta \eta$ serves as a free parameter in the above group detection method. Clearly, if $\Delta \eta$ is too high, some critical wave groups leading to extreme ship motions may be missed. If $\Delta \eta$ is too small, $P_{LA}(l, a)$ may become more complex together with a higher variability of detailed wave elevations for each $(l, a)$, which both negatively impact the convergence of the sequential sampling method. Therefore, a proper value of $\Delta \eta$ is important for the success of the proposed method, but this can only be determined empirically. In section~\ref{sec:validation}, we test values of $\Delta \eta$ around $H_s/2$ and show that such choices provide satisfactory results.  

\begin{figure}[htb!]
  \centering
  \begin{subfigure}{0.5\textwidth}
    \includegraphics[width=\textwidth]{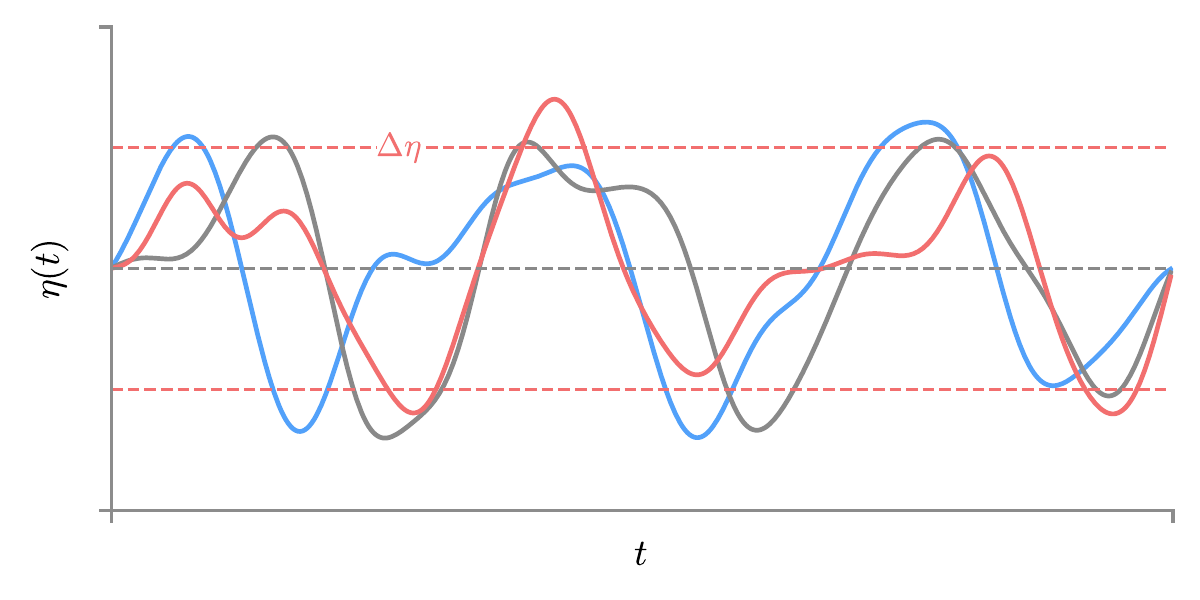}
  \end{subfigure}
  
  \caption{An example of three different forms of wave groups sharing the same group parameters, in this case \( (l, a) = (3T_p, 0.7H_s) \).}
  \label{fig:randomness}
\end{figure}

While the above parameterization method allows description of the wave field in a low-dimensional space, groups with the same $(l, a)$ parameter may have significantly different detailed elevation shapes (see Figure~\ref{fig:randomness}). These different detailed elevations, together with the random initial condition of ship encountering each wave group, lead to a random exceeding time for groups with the same parameters. To account for this randomness, the temporal exceeding probability needs to be formulated as (now dropping the superscript a as this is exact): 
\begin{equation}
P_{\mathrm{temp}}
\;=\;\frac{m}{T_{\mathrm{end}}}
\;\int\;\mathbb{E}_{\omega}\bigl[S(l,a,\omega)\bigr]\;\,P_{LA}(l,a)\,\mathrm{d}l\,\mathrm{d}a,
\label{eq:P_temp_estimate}
\end{equation}
where $S(l,a,\omega)$ with \(\omega\) is a random seed accounting for the randomness mentioned above.

We further note that $\mathbb{E}_{\omega}\bigl[S(l,a,\omega)\bigr]$ in Eq.~\eqref{eq:P_temp_estimate} is quantitatively different from $S(l, a)$ in Eq.~\eqref{eq:ptemp_approx_narrow}. This is because when $S(l, a)$ is approximated by Eq.~\eqref{eq:S_before} with training data $D$, there is no guarantee that each function evaluation provides the mean value over all randomness (which is especially the case for varying ship initial conditions, see \cite{gong2022effects}).
Our next task is to describe the procedure to approximate $S(l, a, \omega)$ through a surrogate GPR model and a sequential sampling method to reduce the number of required samples.

\subsubsection{Surrogate GPR model}
\label{sec:surrogate_GPR}

Construction of a surrogate model for \( S(l, a, \omega) \) is a common procedure in BED. Given a dataset \( D = \{(l^i, a^i, \omega), S^i\}_{i=1}^{n} \) consisting of \( n \) inputs of $(l,a)$ and their exceeding time $S$ as responses, one may naively think that a GPR can be placed on the function $S(l, a, \omega) \equiv \bar{S}(l, a) + \delta(\omega)$, i.e.,

\begin{equation}
    \bar{S}(l, a) | D \sim \text{GP}\left( \mu_{\bar{S}}(l, a | D), \sigma^2_{\bar{S}}(l, a | D) \right),
\end{equation}

\begin{equation}
    \delta(\omega) \sim N (0, \sigma_0^2).
    \label{eq:noise_for_S}
\end{equation}

However, the above GPR approximation is not appropriate due to two reasons. (1) $\delta(\omega)$ does not follow a Gaussian distribution as desired in Eq.~\eqref{eq:noise_for_S}. To understand this point, simply consider a situation that $\bar{S}(l, a)$ is positive and close to zero, so that value of $\delta(\omega)$ is limited (since $S(l, a, \omega)$ has to be non-negative) and cannot be approximated by a Gaussian distribution. (2) $\bar{S}(l, a)$ is not a smooth function (zero for most cases and positive in a small region with a non-smooth transition) which is difficult to be represented as a Gaussian process. We note that reason (2) is also present in the deterministic case, which has been discussed properly in \cite{gong2022efficient}. To address both issues, we define an auxiliary function (i.e., similar procedure as in \cite{gong2022efficient} but here we also include the randomness parameter $\omega$):
\begin{equation}
h(l, a, \omega) = \begin{cases} \dfrac{S(l, a, \omega)}{l}, & \text{if } S(l, a, \omega) > 0 \\ \dfrac{r_{\max}(l, a, \omega) - r_s}{r_s}, & \text{if } S(l, a, \omega) = 0 \end{cases},
\label{eq:h_auxiliary}
\end{equation}
where \( r_\text{max}(l, a, \omega) = \text{max}_t|r(t; l, a, \omega)|\) is the group-maximum response in group \((l, a)\). When \(S=0\), function \(h\) takes negative values with \(r_\text{max}-r_s\) serving as a ‘‘negative penalty’’, quantifying how far the response is from the threshold. We note that by definition $h(l, a, \omega)$ is a much smoother function than $S(l, a, \omega)$ and is rigorously bounded in $[-1,1]$. Therefore, both issues associated with GPR on $S(l, a, \omega)$ are resolved, and we are ready to place a GPR surrogate model on $h(l, a, \omega)$.

Given the dataset \( D = \{(l^i, a^i, \omega), h^i\}_{i=1}^{n} \) and the decomposition $h(l, a, \omega) = \bar{h}(l, a) + \delta(\omega)$, we can place the GPR approximation on $\bar{h}$:
\begin{equation}
\bar{h}(l, a) | D \sim \text{GP}\left( \mu_{\bar{h}}(l, a | D), \sigma^2_{\bar{h}}(l, a | D) \right),
\label{eq:h_gp}
\end{equation}
with the randomness term:
\begin{equation}
    \delta(\omega) \sim N (0, \sigma_0^2).
    \label{eq:noise_for_h}
\end{equation}
where \( \mu_{\bar{h}} \) is the posterior mean and \( \sigma_{\bar{h}}^2 \) is the posterior variance. The randomness \( \delta(\omega) \) is modeled by a zero-mean Gaussian distribution with constant variance \( \sigma_0^2 \), implying a case of homoscedastic randomness. We note that the whole algorithm can be extended to the case of heteroscedastic randomness $\delta(l, a, \omega) \sim N (0, \sigma_0^2(l,a))$ with techniques developed in \cite{gong2022sequential}. However, we find that using homoscedastic randomness is sufficient for the current purpose and we will keep it for the current paper. Once \( h \) is obtained through GPR, we can reconstruct \( S(l, a, \omega) \) by \(S(l, a, \omega) \equiv \mathbf{1}\left( h(l, a, \omega) \right) \text{min}(1, h(l, a, \omega)) l\). The $\text{min}$ operator is because $S$ cannot physically exceed $l$, meaning that the ship
cannot spend more than the entire group length above the threshold.

\subsubsection{Acquisition function}

Given the function \(h\) (and \(S\)) learned from the GPR based on dataset \(D\), \( P_{\text{temp}} \) can be considered as a random variable with its randomness resulting from the uncertainty in \(h\). The uncertainty in \( P_{\text{temp}} \) can therefore be estimated as:

\begin{equation}
U | D = \frac{m}{T_{\text{end}}} \int \left( E_{\omega}[S^+(l, a, \omega) | D] - E_{\omega}[S^-(l, a, \omega) | D] \right) P_{LA}(l, a) \, \mathrm{d}l \, \mathrm{d}a,
\label{eq:uncertainty}
\end{equation}
where $S^\pm(l,a,\omega)|D = \mathbf{1}\{h^\pm(l,a,\omega)|D\}(\text{min}(1, h^\pm(l,a,\omega)|D))l$ with $ h^\pm(l,a,\omega)|D = \Bar{h}^\pm(l,a)|D + \delta(\omega)$ being the upper and lower bound of the defined function $h$ in Eq.~\eqref{eq:h_auxiliary}, and $\Bar{h}^\pm(l,a)|D = \mu_{\Bar{h}}(l,a|D) \pm \sigma_{\Bar{h}}(l,a|D)$ being the upper and lower bounds of $\Bar{h}$ prediction. To analytically compute the expectation \(\mathbb{E}_{\omega}[S^{\pm}(l,a,\omega)\mid D]\), we use the following expression:

\begin{eqnarray}
 \mathbb{E}_{\omega}[S^{\pm}(l,a,\omega)\mid D] &=& l\Bigl[
      \mu_{\text{tr}}^{\pm}(l,a\mid D)\,
      \Bigl(\,
         \Phi\!\Bigl(\frac{1-\Bar{h}^{\pm}(l,a)}{\sigma_0}\Bigr)
       - \Phi\!\Bigl(-\frac{\Bar{h}^{\pm}(l,a)}{\sigma_0}\Bigr)
      \Bigr) \Bigr] \nonumber\\ 
      &+& l\Bigl[
     1 - \Phi\!\Bigl(\frac{1-\Bar{h}^{\pm}(l,a)}{\sigma_0}\Bigr)
   \Bigr]
\label{eq:ESpm}   
\end{eqnarray}
where \(\Phi(\cdot)\) and \(\varphi(\cdot)\) are the standard normal cumulative distribution function (CDF),  and standard normal probability density function (PDF),  respectively, and $\mu^\pm_{\text{tr}}$ is the truncated Gaussian mean given by:

\begin{equation}
\mu_{\text{tr}}^{\pm}(l,a\mid D)
=
\Bar{h}^{\pm}(l,a)
-
\sigma_0
\frac{
   \varphi\!\Bigl(\frac{1-\Bar{h}^{\pm}(l,a)}{\sigma_0}\Bigr)
 - \varphi\!\Bigl(-\frac{\Bar{h}^{\pm}(l,a)}{\sigma_0}\Bigr)
}{
   \Phi\!\Bigl(\frac{1-\Bar{h}^{\pm}(l,a)}{\sigma_0}\Bigr)
 - \Phi\!\Bigl(-\frac{\Bar{h}^{\pm}(l,a)}{\sigma_0}\Bigr)
}.
\label{eq:truncMean}
\end{equation}
The expression in Eq.\eqref{eq:ESpm} has two terms representing contributions from two different intervals of $h$, where the first term accounts for the contribution when $h$ lies between \([0, 1]\), and the second term accounts for the contribution when $h$ exceeds 1. We note that when $h$ is negative, the corresponding $S$ is 0, contributing nothing to the expectation.

Given the uncertainty expressed in Eq.~\eqref{eq:uncertainty}, we then select the next sample which is expected to reduce the uncertainty in $P_{\text{temp}}$ significantly, i.e., a sample corresponding to the maximum of the integrand in Eq.\eqref{eq:uncertainty}:

\begin{equation}
l^*, a^* = \arg \max_{l,a} 
\left( E_\omega[S^+(l,a,\omega)| D] - E_\omega[S^-(l,a,\omega)| D] \right)  P_{LA}(l,a).
\label{enq:argmax}
\end{equation}

At each iteration, the acquisition function is optimized to identify the most informative point \((l^*, a^*)\) within the parameter space. We then randomly locate a wave group in the original wave field with parameters \((l^*, a^*)\) (or sufficiently close to it) to capture the randomness in detailed wave shape. In addition, we run the next ship motion simulation from $\num{1}~T_p$ ahead of the \((l^*, a^*)\) group to naturally capture the randomness in ship initial condition encountering the \((l^*, a^*)\) group (see \cite{gong2022effects} for an extensive validation of this approach). Additionally, the simulation is continued for approximately $\num{1}~T_p$ after the group to account for the time-delayed responses. This process is then continued such that for each iteration, the sequential sample is placed at a location that contributes most significantly to reducing the uncertainty in \( P_{\text{temp}} \). The process ends with the computation of \( P_{\text{temp}} \) using Eq.~\eqref{eq:P_temp_estimate}. We summarize the full algorithm in Algorithm~\ref{alg:bed_temp}, with the code available on Github\footnote{https://github.com/umbrellagong/gpbroad}.

\begin{algorithm}[htb!]
\caption{Bayesian Experimental Design for Temporal Exceeding Probability}
\label{alg:bed_temp}
\begin{algorithmic}[1]
    \REQUIRE Number of initial samples \( n_{\text{init}} \), number of sequential samples \( n_{\text{seq}} \)
    \STATE \textbf{Initialize:} Dataset \( D = \{ (l^i, a^i, \omega), h^i \}_{i=1}^{n_{\text{init}}} \)
    \FOR{ \( j = 1 \) to \( n_{\text{seq}} \) }
        \STATE Train the surrogate model using GPR with Eq.~\eqref{eq:h_gp} and Eq.~\eqref{eq:noise_for_h}
        \STATE Compute the uncertainty \( U | D \) using Eq.~\eqref{eq:uncertainty}
        \STATE Select the next sample \( (l^{j+1}, a^{j+1}) \) using Eq.~\eqref{enq:argmax}
        \STATE Simulate \( h^{j+1} = h(l^{j+1}, a^{j+1}, \omega) \)
        \STATE Update dataset \( D = D \cup \{ (l^{j+1}, a^{j+1}), h^{j+1} \} \)
    \ENDFOR
    \STATE \textbf{Output:} Compute \( P_{\text{temp}} \) using the surrogate model and Eq.~\eqref{eq:P_temp_estimate}
\end{algorithmic}
\end{algorithm}

\section{Validation of the method}
\label{sec:validation}
In this section, we present the results of the proposed method applied to a wide range of cases with varying wave spectral bandwidth, nonlinear ship motion model parameters, and exceeding thresholds. The results of $P_{\text{temp}}$ are compared to their exact values $P^e_\text{temp}$ computed using Eq.~\eqref{eq:P_temp} from simulating the ship motion model with sufficiently long time series of wave excitation. In addition, we compare the performance of the developed sequential sampling approach with other commonly used methods such as random and LH sampling to show the superiority of the developed approach. 

\subsection{Case setup}
To set up our test cases, we consider the nonlinear roll motion of a ship subject to a wave field with a broadband spectrum. Specifically, we use a JONSWAP spectrum with spectral peak period $T_p=\SI{15}{s}$, significant wave height $H_s=\SI{12}{m}$, and a peak enhancement factor $\gamma$ with two chosen values of $\gamma$ =1 and 3 (both realistically broadband). The ship motion is assumed to be modeled by a nonlinear roll equation, which models the ship roll response $r(t)$ due to nonlinear resonance and parametric roll in oblique irregular waves with time series of elevation \(\eta(t)\):

\begin{equation} 
\ddot{r} + \alpha_1 \dot{r} + \alpha_2 \dot{r} |\dot{r}| + \left( \beta_1 + \varepsilon_1 \cos(\theta) \eta(t) \right) r + \beta_2 r^3 = \varepsilon_2 \sin(\theta) \eta(t), 
\label{eq:roll_eq} 
\end{equation}
where \( \alpha_1 \) and \( \alpha_2 \) are linear and nonlinear damping coefficients, respectively; \( \beta_1 \) and \( \beta_2 \) are linear and nonlinear restoring coefficients; \( \varepsilon_1 \) and \( \varepsilon_2 \) are parametric and direct wave excitation coefficients; and \( \theta \) is the encounter angle between the ship and the wave. These empirical coefficients are set as \( \alpha_1 = 0.35 \), \( \alpha_2 = 0.06 \), \( \beta_1 = 0.04 \), \( \theta = \pi/6 \), and \( \varepsilon_2 = 0.012 \), as in \cite{gong2022efficient}. We vary the other two parameters, namely $(\beta_2, \varepsilon_1) \in \{ (-0.1, 0.008), \, (-0.2, 0.008), \, (-0.1, 0.016) \}$ to represent different levels of nonlinear restoring and parametric excitations in the ship roll dynamics. We note that the simple model in Eq.~\eqref{eq:roll_eq} is used since we need the exact result $P^e_\text{temp}$ for validation, but our method is not restricted to a particular choice of ship model. For example, one can easily replace Eq.~\eqref{eq:roll_eq} by a CFD calculation as demonstrated earlier in \cite{gong2022efficient}.

In summary, the parameter values in all tested cases are listed in Table~\ref{tab:case_study_params}, which also includes different values of $\Delta \eta$, chosen as values close to $H_s/2$ (in particular, with three choices: $H_s$ and $H_s/2\pm 1$) to benchmark the sensitivity of the method to its free parameter as discussed in section~\ref{sec:method}.

\begin{table}[htb!]
  \centering
  \begin{threeparttable}
    \caption{Parameter values explored in the case studies for validating the proposed method.}
    \label{tab:case_study_params}
    \begin{tabular}{ll}
      \toprule
      \textbf{Parameter} & \textbf{Values} \\
      \midrule
      \makecell[l]{Peak enhancement factor \\ $\gamma$ (–)} & 1, 3 \\
      \makecell[l]{Ship-motion dynamical model \\ $(\beta_2,\;\varepsilon_1)$ ($\si{s^{-2}\radian^{-2}},\;\si{m^{-1}s^{-2}}$)\tnote{*}} & (–0.1, 0.008), (–0.2, 0.008), (–0.1, 0.016) \\
      \makecell[l]{Exceeding threshold \\ $r_s$ (\si{\radian})} & 0.30, 0.35 \\
      \makecell[l]{Wave-parametrization threshold \\ $\Delta\eta$ (\si{m})} & 5, 6, 7 \\
      \bottomrule
    \end{tabular}
    \begin{tablenotes}
      \footnotesize
      \item[*] Hereafter, the units of $\beta_2$ and $\varepsilon_1$ are omitted for conciseness.
    \end{tablenotes}
  \end{threeparttable}
\end{table}

\subsection{Results}
\begin{figure}[htb!]
  \centering
  \begin{subfigure}{0.8\textwidth}
    \includegraphics[trim={0cm 0cm 0cm 0cm}, width=\textwidth]{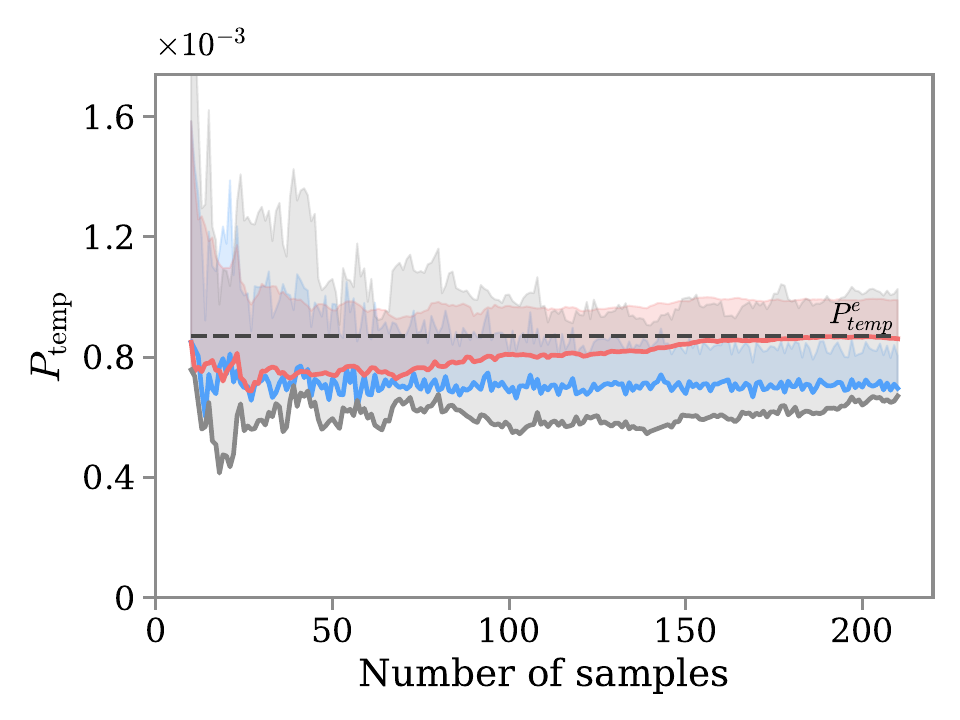}
  \end{subfigure}
    
    \definecolor{LHColor}{rgb}{0.32, 0.63, 0.98}
    \definecolor{SeqColor}{rgb}{0.95, 0.44, 0.44}
    \definecolor{RandColor}{rgb}{0.54, 0.54, 0.54}
    \definecolor{PtempColor}{rgb}{0, 0, 0}
    
     \caption{Temporal exceeding probability $P_{\text{temp}}$ for the case with $\varepsilon_1 = 0.008$, $\beta_2 = -0.2$, $\gamma = 3$, $r_s = \SI{0.35}{\radian}$, and $\Delta \eta = \SI{5}{m}$, computed from LH sampling (\textcolor{LHColor}{\rule[0.5ex]{1.5em}{1pt}}), sequential sampling (\textcolor{SeqColor}{\rule[0.5ex]{1.5em}{1pt}}), random sampling (\textcolor{RandColor}{\rule[0.5ex]{1.5em}{1pt}}), and its exact reference value $P^e_{\text{temp}} = \num{0.00087}$ (\textcolor{RandColor}{\rule[0.5ex]{0.25em}{1pt} \hspace{-0.5em} \rule[0.5ex]{0.25em}{1pt} \hspace{-0.5em} \rule[0.5ex]{0.25em}{1pt}}).}
    \label{fig:P_temp}
\end{figure}

We start by showing a typical result for the case with $\varepsilon_1 = 0.008$, $\beta_2 = -0.2$, $\gamma = 3$, $r_s = \SI{0.35}{\radian}$, and $\Delta \eta = \SI{5}{m}$.  Figure~\ref{fig:P_temp} plots the computed $P_{\text{temp}}$ as a function of sampling numbers using sequential sampling, random sampling, and LH sampling, together with its exact value $P^e_\text{temp}$. All the sampling results are obtained as an ensemble of \num{100} repeated experiments, i.e., we repeat the respective sampling calculations \num{100} times, each time starting from \num{10} different initial LH samples. The results are therefore presented as the mean plus one standard deviation shown on the upper side of the mean curve, to improve the fairness of the comparison. We see in Figure~\ref{fig:P_temp} that results from sequential sampling successfully approaches $P_{\text{temp}}^e$ in less than \num{200} samples, while the results from random and LH samplings fail to capture $P_{\text{temp}}^e$ even at the end of \num{210} samples. In addition, the results from sequential and LH samplings are associated with similar uncertainty  (or standard deviation), both smaller compared to that of random sampling. This is a point which will be further discussed later.

\begin{figure}[htb!]
  \centering
  \begin{subfigure}{0.8\textwidth}
    \includegraphics[trim={0cm 0cm 0cm 0cm}, width=\textwidth]{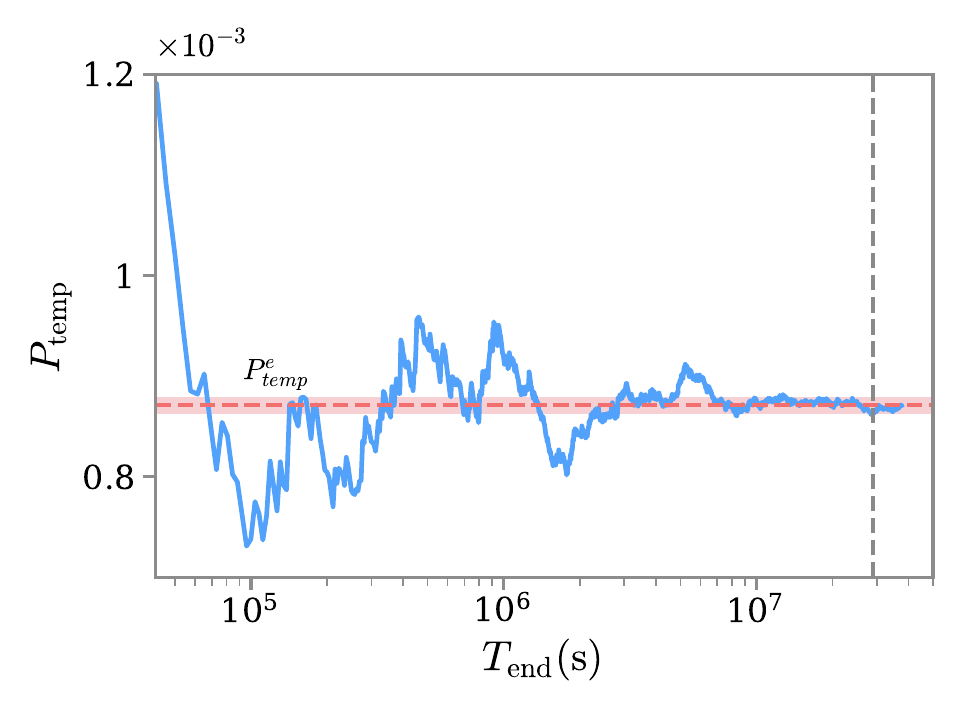}
  \end{subfigure}
    
    \definecolor{BLUE}{rgb}{0.32, 0.63, 0.98}
    \definecolor{PINK}{rgb}{0.95, 0.44, 0.44}
    \definecolor{GRAY}{rgb}{0.54, 0.54, 0.54}
    \definecolor{BLACK}{rgb}{0, 0, 0}
    
    \caption{Brute-force computation of $P_\text{temp}$ using Eq.~\eqref{eq:P_temp} with different $T_\text{end}$ (i.e., length of time series to be simulated) for the case $\varepsilon_1 = 0.008$, $\beta_2 = -0.2$, $\gamma = 3$, threshold $r_s = \SI{0.35}{\radian}$. Exact value of $P_\text{temp}^e$, calculated via $T_\text{end}=\num{2.56e6}~T_p$ is indicated by (\textcolor{PINK}{\rule[0.5ex]{0.3em}{1pt} \rule[0.5ex]{0.3em}{1pt} \rule[0.5ex]{0.3em}{1pt}}), with its $\pm \SI{1}{\percent}$ bound indicated by the shaded region. The vertical dashed (\textcolor{GRAY}{\rule[0.5ex]{0.3em}{1pt} \rule[0.5ex]{0.3em}{1pt} \rule[0.5ex]{0.3em}{1pt}}) line marks the value of $T_\text{end}\approx \SI{3e7}{s}$ beyond which the calculated $P_\text{temp}$ lies consistently within the $\SI{1}{\percent}$ bound.}
    \label{fig:1percent_error_truth}
\end{figure}

We further characterize the computational cost saving by our method. In the result shown in Figure~\ref{fig:P_temp}, it takes about \num{182} samples for our sequential sampling method to reach within \SI{1}{\percent} error bar around $P^e_\text{temp}$. Considering each sequential sample contain a simulation of time series of \SI{68.8}{s} on average, the total simulation time with sequential sampling method is about \SI{12520.4}{s}. In contrast, the exact calculation of $P_{\text{temp}}^e$ requires \SI{28876800}{s} simulation for the obtained value to reach \SI{1}{\percent} accuracy, as shown in Figure~\ref{fig:1percent_error_truth}. Therefore, by employing our approach equipped by the new wave group parameterization and sequential sampling, we are expected to save approximately \num{2300} times computational resources.

Our next goal is to benchmark the performance of our developed method with all \num{36} cases listed in Table~\ref{tab:case_study_params}. For this purpose, we first introduce two metrics to characterize the performance of a method. The first one is the Normalized Mean Absolute Error (NMAE) to describe the relative error of an estimation normalized by the true value, defined as:
\begin{equation}
    \text{NMAE} = \frac{1}{N} \sum_{i=1}^{N} \left| \frac{\hat{P}^i_{\text{temp}} - P_{\text{temp}}^e}{P_{\text{temp}}^e} \right|,
    \label{eq}
\end{equation}
where $\hat{P}_{\text{temp}}^i$ is the estimated $P_{\text{temp}}$ using a particular sampling method (from random, LH, and sequential samplings) at the $i$-th trial, and $N$ is the total number of trials. To be consistent with results in Figure~\ref{fig:P_temp}, we will keep using $\hat{P}_{\text{temp}}^i$ obtained at 210 samples and $N=100$. The second metric is the standard deviation of the estimated $P_{\text{temp}}$  normalized by the true $P_{\text{temp}}^e$ value (NSTD), also evaluated at \num{210} samples with respect to the $N=100$ ensembles.

\begin{figure}
    \centering
    \includegraphics[width=0.9\linewidth]{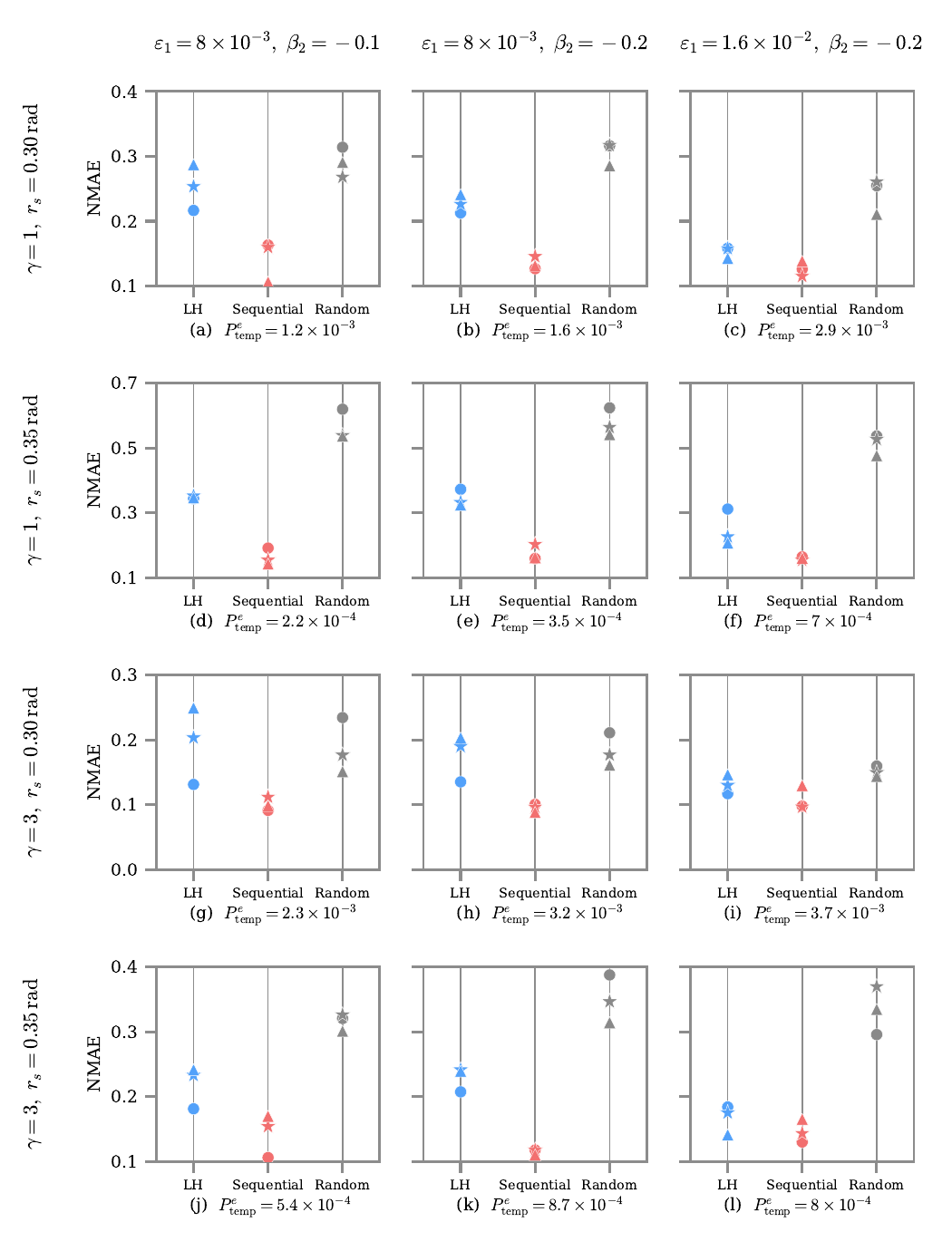}
    \caption{Values of NMAE for the \num{36} test cases with sequential, random, and LH samplings. Each column is for one particular dynamical system: first column (a, d, g, j) for $\varepsilon_1=0.008$, $\beta_2=-0.1$; second column (b, e, h, k) for $\varepsilon_1=0.008$, $\beta_2=-0.2$; and third column (c,  f, i, l) for $\varepsilon_1=0.016$, $\beta_2=-0.2$. Each row is for one choice of spectral bandwidth $\gamma$ and exceeding threshold $r_s$: first row (a, b, c) for $\gamma = 1$, $r_s = \SI{0.30}{\radian}$; second row (d, e, f) for $\gamma = 1$, $r_s = \SI{0.35}{\radian}$; third row (g, h, i) for $\gamma = 3$, $r_s = \SI{0.30}{\radian}$; and fourth row (j, k, l) for $\gamma = 3$, $r_s = \SI{0.35}{\radian}$. Different symbols in each figure represent results with $\Delta \eta=\SI{5}{m}$ (\textcolor{gray}{\boldmath$\largecircle$}), $\Delta \eta=\SI{6}{m}$ (\textcolor{gray}{\boldmath$\largestar$}) and $\Delta \eta=\SI{7}{m}$ (\textcolor{gray}{\boldmath$\largetriangleup$}). Value of $P^e_\text{temp}$ is listed for each figure.}
    \label{fig:errors}
\end{figure}

\begin{figure}
    \centering
    \includegraphics[width=0.9\linewidth]{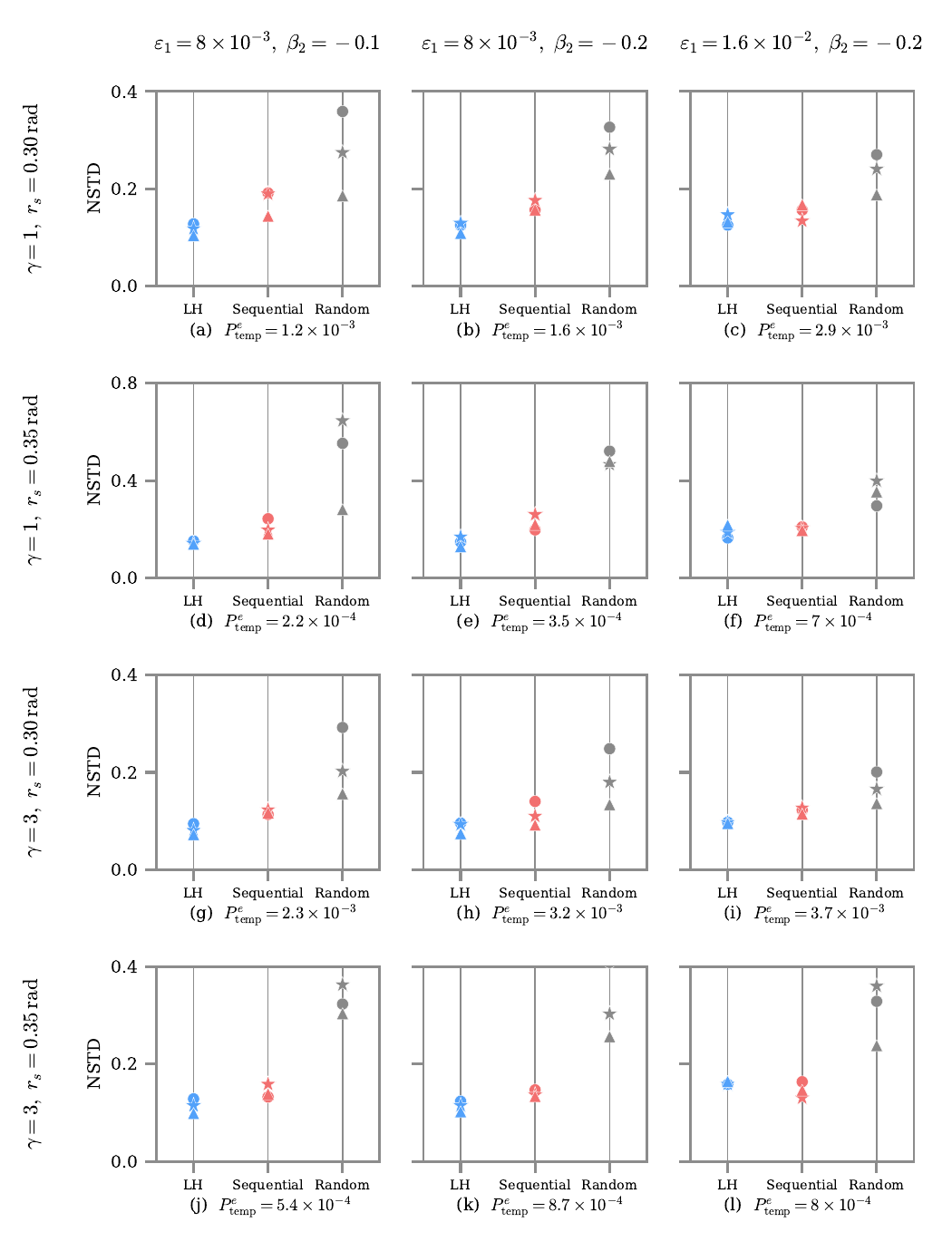}
    \caption{Values of NSTD for the \num{36} test cases with sequential, random, and LH samplings. Each column is for one particular dynamical system: first column (a, d, g, j) for $\varepsilon_1=0.008$, $\beta_2=-0.1$; second column (b, e, h, k) for $\varepsilon_1=0.008$, $\beta_2=-0.2$; and third column (c,  f, i, l) for $\varepsilon_1=0.016$, $\beta_2=-0.2$. Each row is for one choice of spectral bandwidth $\gamma$ and exceeding threshold $r_s$: first row (a, b, c) for $\gamma = 1$, $r_s = \SI{0.30}{\radian}$; second row (d, e, f) for $\gamma = 1$, $r_s = \SI{0.35}{\radian}$; third row (g, h, i) for $\gamma = 3$, $r_s = \SI{0.30}{\radian}$; and fourth row (j, k, l) for $\gamma = 3$, $r_s = \SI{0.35}{\radian}$. Different symbols in each figure represent results with $\Delta \eta=\SI{5}{m}$ (\textcolor{gray}{\boldmath$\largecircle$}), $\Delta \eta=\SI{6}{m}$ (\textcolor{gray}{\boldmath$\largestar$}) and $\Delta \eta=\SI{7}{m}$ (\textcolor{gray}{\boldmath$\largetriangleup$}). Value of $P^e_\text{temp}$ is listed for each figure.}
    \label{fig:errors_std}
\end{figure}

The values of NMAE and NSTD are plotted in Figures~\ref{fig:errors} and \ref{fig:errors_std}, respectively, for the 36 test cases. In Figure~\ref{fig:errors} we see that NMAE resulted from our sequential sampling method is consistently the smallest among all three tested sampling methods. For most cases, NMAEs from sequential sampling are below or around \SI{15}{\percent}, except that for cases with $\gamma=1$ and $r_s=\SI{0.35}{\radian}$, NMAE can rise up to \SI{20}{\percent} in Figures~\ref{fig:errors}d and \ref{fig:errors}e (considering the $P_{\text{temp}}^e$ for these cases is O($\num{e-4}$), this error should still be considered small). This is natural because these cases are associated with the larger spectral bandwidth and higher exceeding threshold, where the former factor leads to more irregular wave patterns (thus groups) and the latter leads to rarer exceedance. For all cases (including those of Figures~\ref{fig:errors}d and \ref{fig:errors}e), we have verified that if we further increase the sequential sampling numbers, the result may further improve, but the final converged results may be limited by the slight inaccuracies inherent in the group parameterization method. In addition, from all tested cases in Figure~\ref{fig:errors}, we find no consistent preference on the value of $\Delta \eta$. Instead, from the results we argue that all choices of $\Delta \eta$ around $H_s/2$ provide reasonable and close estimation for $P_{\text{temp}}$.

The values of NSTD plotted in Figure~\ref{fig:errors_std} show a somewhat different trend. In general, the results of LH sampling achieves the smallest value of NSTD among the three sampling methods. This is because LH sampling ensures samples to be uniformly distributed in the $(L,A)$ space with less variability over trials. However, practically the lower NSTD of LH sampling method is not a favorable situation, since the method is meanwhile associated with larger NMAE. Such a situation of lower NSTD and higher NMAE represents a bias in the estimation of $P_{\text{temp}}$ by LH sampling. The sequential sampling, on the other hand, is associated with slightly larger NSTD but significantly smaller NMAE, therefore serving as a much more balanced method for computing $P_{\text{temp}}$.

\begin{figure}[htb!]
    \centering
    \begin{subfigure}{0.4\textwidth}
        \includegraphics[width=\textwidth]{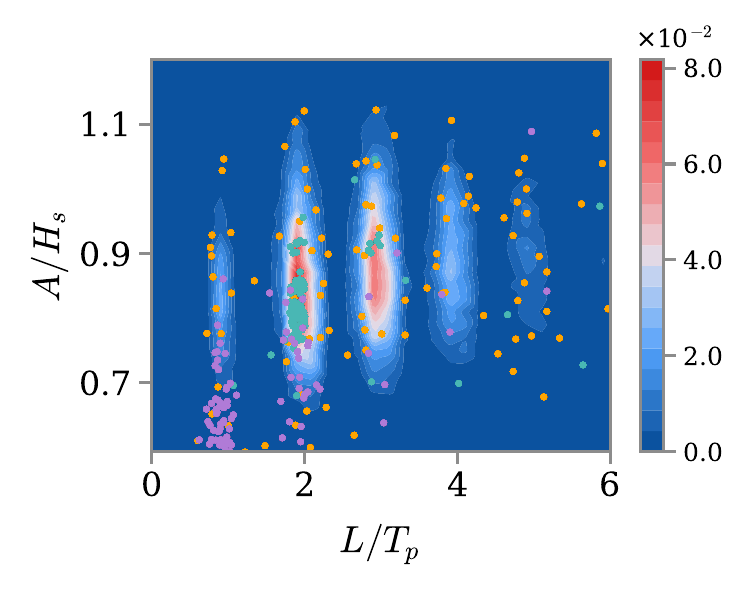}
        \vspace{-2em}
        \caption{}
    \end{subfigure}
    \begin{subfigure}{0.4\textwidth}
        \includegraphics[width=\textwidth]{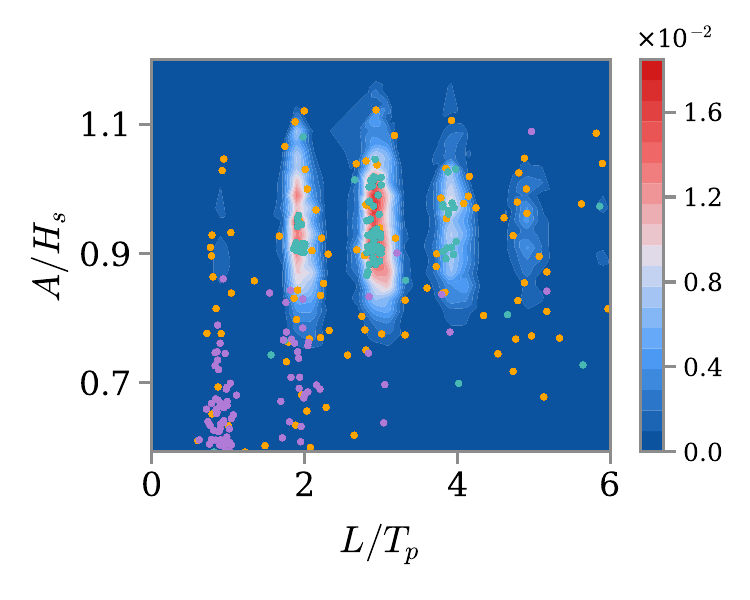}
        \vspace{-2em}
        \caption{}
    \end{subfigure}
    
    \definecolor{random_color}{rgb}{0.69, 0.48, 0.84}
    \definecolor{lh_color}{rgb}{1.00, 0.65, 0.00}
    \definecolor{sequential_color}{rgb}{0.29, 0.72, 0.71}
    
    \caption{Sample locations from sequential (\textcolor{sequential_color}{$\bullet$}), random (\textcolor{random_color}{$\bullet$}), and LH (\textcolor{lh_color}{$\bullet$}) samplings for the case with parameters $\varepsilon_1 = 0.008$, $\beta_2 = -0.1$, $\gamma = 3$, $\Delta \eta = \SI{7}{m}$, and varying $r_s=\SI{0.30}{\radian}$ and $r_s = \SI{0.35}{\radian}$ for (a) and (b). Contour plots of $E_{\omega}[S(l, a, \omega)] P_{LA}(l, a)$ are included in both sub-figures.}
    \label{fig:samples}
\end{figure}

We finally explore the reason underlying the high efficiency of sequential sampling in computing $P_{\text{temp}}$. Figure~\ref{fig:samples} shows the sampling locations of the three sampling methods in the space of $(L,A)$. We also plot the contour of the function $E_{\omega}[S(l, a, \omega)] P_{LA}(l, a)$ in the figure, which illustrates the importance of sampling location to the result of $P_{\text{temp}}$ according to Eq.~\eqref{eq:P_temp_estimate}. Preferably, we would like regions with higher $E_{\omega}[S(l, a, \omega)] P_{LA}(l, a)$ to be sampled more so that the regions contributing dominantly to $P_{\text{temp}}$ is approximated more accurately in GPR. This is indeed what happens in the sequential sampling, where we see most sample points are concentrated near the peaks of $E_{\omega}[S(l, a, \omega)] P_{LA}(l, a)$ that vary in different cases. In contrast, random sampling and LH sampling lack this feature, with the former concentrating on regions with higher $P_{LA}(l, a)$ and the latter uniformly distributed in the $(L,A)$ space. Therefore, the favorable efficiency of sequential sampling to compute $P_{\text{temp}}$ can be expected from this visualization of sampling locations. 

\section{Conclusion}

In this paper, we continue to develop our efficient method in evaluating the temporal exceeding probability \(P_{\text{temp}}\) of ship responses subject to an incoming wave field, i.e., the fraction of time that the response exceeds a specified threshold. The current work improves upon our previous paper \cite{gong2022efficient} in two aspects. (1) It extends the calculation into wave fields of general spectra, including the realistically broadband cases. (2) It captures the variability of response for given group parameters $l$ and $a$, which originates from the variable wave form and ship encountering initial conditions. Both improved features are integrated into our general framework including wave group parameterization and sequential sampling enabled by GPR and a newly-developed acquisition function. We test the performance of our new framework in many cases with different ship motion models, wave spectral bandwidth and exceeding thresholds. We show that in all cases our method achieves favorable results, with an error of O(\SI{15}{\percent}) or below in most of the cases and a saving of computational cost of \num{2300} times compared to the brute-force computation. 

\section*{ACKNOWLEDGEMENT}
This research is supported by the Office of Naval Research, USA grant N00014-24-1-2266.

\appendix
\section{Gaussian process regression}
\label{appendix:first}

We consider the task of inferring the input-to-response (ItR) function from a dataset \( D = \{X^i, y^i\}_{i=1}^n \) consisting of \( n \) inputs \( \boldsymbol{X} = \{X^i \in \mathbb{R}^d\}_{i=1}^n \) (e.g.,\ \( X=(l,a)\) in section~\ref{sec:surrogate_GPR}) and the corresponding outputs \( \boldsymbol{y} = \{y^i \in \mathbb{R}\}_{i=1}^n \) (e.g.,\ \(y=h\) in section~\ref{sec:surrogate_GPR}).

GPR assumes the observed output equals the latent function \( f(X) \) plus i.i.d. Gaussian noise with constant variance  \( \gamma_0^2 \) at all $X$:
\begin{equation}
y = f(X) + R, \quad R \sim N(0, \gamma_0^2). \tag{A.1}
\end{equation}

A prior, representing our beliefs over all possible functions we expect to observe, is placed on \( f \) as a Gaussian process \( f(X) \sim \text{GP}(0, k_f(X, X')) \) with zero mean and covariance function \( k_f \). Following the Bayes' theorem, the posterior prediction for \( f \) given the dataset \( D \) can be derived to be another Gaussian:

\begin{equation}
p(f(X) | D) = \frac{p(f(X), \boldsymbol{y})}{p(\boldsymbol{y})} = \text{GP} (\mu_f(X), \operatorname{cov}_f(X, X')), \tag{A.2}
\end{equation}
with analytically tractable mean \( \mu_f(X) \) and covariance \( \operatorname{cov}_f(X, X') \):
\begin{equation}
\mu_f(X) = k_f(X, \boldsymbol{X})^\top \left( K_f(\boldsymbol{X}, \boldsymbol{X}) + \gamma_0^2 I \right)^{-1} \boldsymbol{y}, \tag{A.3}
\end{equation}
\begin{equation}
\operatorname{cov}_f(X, X') = k_f(X, X') - k_f(X, \boldsymbol{X})^\top \left( K_f(\boldsymbol{X}, \boldsymbol{X}) + \gamma_0^2 I \right)^{-1} k_f(\boldsymbol{X}, X'), \tag{A.4}
\end{equation}
where \( K_f(\boldsymbol{X}, \boldsymbol{X}) \) is the covariance matrix with elements \( [K_f(\boldsymbol{X}, \boldsymbol{X})]_{ij} = k_f(X_i, X_j) \), and \( I \) is the identity matrix. 

For the covariance function $k_f$, we use the Matérn kernel  with smoothness parameter \( \nu = 1.5 \) defined as:
\begin{equation}
k_f(X, X') = \tau^2 \frac{1}{\Gamma(\nu) 2^{\nu - 1}} \left( \sqrt{2\nu} \, \text{dist}(X, X') \right)^\nu K_\nu\left( \sqrt{2\nu} \, \text{dist}(X, X') \right). \tag{A.5}
\end{equation}
Here, \( K_\nu \) is the modified Bessel function of the second kind, \( \Gamma(\nu) \) is the gamma function, and the distance function \( \text{dist}(X, X') \) is computed by:
\begin{equation}
\text{dist}(X, X') = \left( (X - X')^\top \Lambda^{-1} (X - X') \right)^{1/2}, \tag{A.6}
\end{equation}
where \( \tau \) and diagonal matrix \( \Lambda \) are the kernel hyperparameters representing the characteristic amplitude and length scales. The GPR hyperparameters \( \theta = \{\tau, \Lambda, \gamma_0\} \) are determined by maximizing the likelihood function \( p(D|\theta) \equiv p(\boldsymbol{y}|\theta) = N(0, K_f(\boldsymbol{X}, \boldsymbol{X}) + \gamma_0^2 I) \).

\bibliographystyle{elsarticle-harv}

\end{document}